\documentclass[epj,referee]{svjour}
\usepackage{amsmath}
\usepackage{graphics}
\begin{document}
\title{Dependence of the average to-node distance on the node degree for random graphs and growing networks}
\author{K.Malarz \and K.Ku{\l}akowski 
} 
%
%
\institute{
Faculty of Physics and Applied Computer Science,
AGH University of Science and Technology,
al. Mickiewicza 30, PL-30059 Krak\'ow, Poland.
\email{malarz@agh.edu.pl, kulakowski@novell.ftj.agh.edu.pl}
}
\date{Received: date / Revised version: date}
%
\abstract{
In a connected graph, nodes can be characterised locally (with their degree $k$) or
globally (e.g. with their average length path $\xi$ to other nodes).
Here we investigate how $\xi$ depends on $k$.
The numerical algorithm based on the construction of the distance matrix is applied to random graphs and the growing networks: the scale-free ones and the exponential ones.
The results are relevant for
search strategies in different networks.  
\PACS{
{02.10.Ox}{Combinatorics; graph theory}  
\and
{05.10.-a}{Computational methods in statistical physics and nonlinear dynamics}
} 
\keywords{evolving networks -- simple graphs and trees -- random graphs -- numerical calculations -- sociophysics}
} 
\authorrunning{K.Malarz, K.Ku{\l}akowski}
\titlerunning{Degree and average path length}
\maketitle
%

\section{Introduction}
Recent interest in analytical and numerical research of growing networks
\cite{ab,drm,nwm} was initiated by a seminal paper of Barab\'asi and
Albert \cite{ba}. The authors demonstrated that a natural algorithm of
growing produces a scale-free power law distribution of the node degree,
i.e. of the number of edges of a node. Moreover, this power law has been
found to appear in several existing networks, as the actor collaboration
network, the WWW, and the power grid network \cite{ba}. Now, many other
examples of this universal pattern has been discovered \cite{nwm}, and the
list seems still open. The idea of a growing network emerges as a new paradigm
of interdisciplinary importance.

The growing process is understood as a successive adding of new nodes, each linked
to older ones by $m$ edges. When $m=1$ a so-called {\em tree} appears.
A tree is a connected graph without cyclic paths. For $m\ge 2$
--- when a newly attached node is linked to more than one node --- cyclic paths
are possible and the formed structure is termed as {\em a simple graph}
\cite{clark,wilson,drmesa}.
While the network --- a tree or a simple graph --- grows,
existing nodes to which the new ones are linked can be selected preferentially,
i.e. with the probability proportional to their degree.
In this case, the degree distribution
is given by the power law, i.e. $P(k)\propto k^{-\gamma}$.
If the nodes are selected randomly, the degree distribution is exponential,
i.e. $P(k)=2^{-k}$.

One of the striking features of many growing networks is the small-world effect
\cite{ws}.
Namely, in such networks the mean distance $d$ between nodes increases with the number $N$ of nodes only as $\ln(N)$ or slower.
For example, the actor collaboration network is formed of 449 thousand
nodes; two actors are linked if they happened to play roles in the same movie.
The mean distance $d$, i.e. the mean number of links between actors, is
less than 3.5 \cite{nwm}.

Actually, the small-world effect in human relations has been discovered more
than 35 years ago in a brilliant sociometric experiment \cite{mil,tra}.
A group of individuals was asked to send a letter to a target person in
Boston via an acquaintance who was supposed to be closer to the target
than the sender.
The mean length of the letter chain was less than seven.
This experiment was repeated several times \cite{fro}, and it is currently
being continued at Columbia University \cite{col}.
Recently, such considerations happened to inspire a hierarchical model of a
social network \cite{wdn}, where a contact between different groups within
a given hierarchy is possible only via a person who is higher in the hierarchy.

In this kind of contact experiment, to find an appropriate next person in
the path is a nontrivial task, and several strategies are possible \cite{ada,kim}.
One of the most obvious is to find a person most connected, i.e. a neighbouring
node with the highest degree. This strategy has been shown to be effective in
networks with power-law degree distribution, but not in random graphs \cite{ada}.
We note here that all strategies must be ceased once the desired
target is in a reasonably short distance. The discussion below is conducted
with this condition in mind.

In this paper, the problem addressed is if this strategy is effective in
the exponential networks. However, our numerical method is different from the
approach applied in Ref. \cite{ada,kim}.
Here we construct the distance matrix for a given network.
For each node $i$ having the degree $k$, we calculate the mean distance $\xi_i$ to all other nodes in
the network. For a given kind of network (say, scale-free networks) we
calculate the average of $\xi_i$ over all nodes with given degree $k$.
In this way we get a curve $\xi(k)$.
The average distance $d$ can be obtained by averaging $\xi(k)$ over $k$.
It is obvious that $\xi$ decreases with $k$, because on average, the paths from more connected nodes are shorter than the paths from a node with one or two edges only.
If this decrease is sharp, the strategy of the most connected neighbour (MCNS)
is effective, because the path from the selected neighbour to other nodes
is shorter on average.

It is worth mentioning here, that $1/\xi_i$ is a direct measure of the so-called closeness centrality (CC) for a given node \cite{sabidussi}.
A node with high CC is obviously in a good position to get other nodes on short paths.
The MCNS strategy (termed as MAX in Ref. \cite{kim}) is just to increase the node degree.
The slope of the curve $\xi(k)$ then brings information on how this strategy is effective for a given network.
The effectiveness of MCNS for nodes of given $k$ can be evaluated by an index
\begin{equation}
\label{eq-eta}
\eta=-\frac{\partial\xi}{\partial\ln k}.
\end{equation}

In principle, the total effectiveness for a given kind of network should be calculated as an average over all nodes.
In such an average, the majority of nodes have a low degree.
Then, what is relevant is the value of $\eta$ for low $k$.
Instead of averaging, we show that curve $\eta(k)$ carries all the important information.

In the next section, we describe our method of simulation. Later on we show
the results for the scale-free networks, the exponential networks and connected Erd\"os--R\'enyi random graphs (CRG) \cite{ab,crg,crg2}.
The section is closed by a discussion.

\section{Calculations}
A standard way of calculating distances between two nodes is the breadth-first search algorithm \cite{clr,amo}.
Our numerical approach is based on the construction of the distance matrix
$\mathbf{S}$, an element of which $s(i,j)$ indicates the length of the shortest path between nodes $i$ and $j$.
The matrix $\mathbf{S}$ is formed simultaneously with the network growth \cite{mckk,mk,mkkk}.

For the exponential networks, the nodes to which new nodes are attached are selected randomly.
For the scale-free networks, these nodes are selected preferentially, i.e. with the probabilities proportional to their degree \cite{ba}.

For the growing networks, the starting point of
the simulation is a matrix 
\[
\mathbf{S}=\begin{pmatrix} 0 & 1 \\ 1 & 0 \\ \end{pmatrix}
\]
representing only two nodes linked together.  
The subsequent stages of the construction of the matrix $\mathbf{S}$ for
growing trees ($m=1$) and growing simple graphs ($m=2$)
were described in Refs. \cite{mckk,mk,mkkk}.
Here, we present a similar algorithm for the construction of the distance
matrix $\mathbf{S}$ for Erd\"os--R\'enyi CRG \cite{crg,crg2}.

We start the simulation with an $N\times N$ matrix with all non-diagonal
elements equal to $N$, which is larger than the largest possible
distance between any of $N$ connected nodes.
Then --- following the definition of CRG --- we try to link each
node pair randomly with a given probability $p$.
Strictly speaking, we go through all non-diagonal elements of $\mathbf{S}$
and set $s(i,j<i)$ equal to one with the probability $p$.
Obviously, the matrix $\mathbf{S}$ is kept symmetric.
Each time, when a new edge is added, we have to
rebuild the whole matrix $\mathbf{S}$ due to the link between nodes $i$ and $j$:
\begin{equation}
\begin{split}
\label{eq_crg}
\forall 1\le m,n\le N: s(m,n)=
\min \big( s(m,n),\\ s(m,i)+1+s(j,n), s(m,j)+1+s(i,n) \big).
\end{split}
\end{equation}
After such a procedure the matrix $\mathbf{S}_{N\times N}$ contains elements
equal to $N$ only if the graph is not connected.

One could ask if the order of updating the matrix elements could change the final result. Our answer is no, and the argument is as follows. Adding an edge, say
$(m,n)$, we have to check for each pair $(i,j)$ the minimum of the following:
$s(i,j)$ before adding a new edge $(m,n)$, which does not contain this edge by
definition; $s(i,m)+1+s(n,j)$ and $s(i,n)+1+s(m,j)$. The path $s(m,n)$ is
represented above as a unit. No other part $s$ of the path selected as the
minimal one contains the edge $(m,n)$; if the path does contain it, it
contains it twice and therefore it is not minimal. In other words,
there are two possibilities: either the minimal path $(i,j)$ does not
contain the new edge $(m,n)$, or it contains it once. Then, all paths
$s$ used in Eq. \eqref{eq_crg} and selected as minimal do not contain the new edge. Then,
they are not changed by adding this new edge. Therefore the order of updating
these parts is not relevant.

For a given matrix $\mathbf{S}$ we calculate the distribution of node degree
$P(k)$ and the average distance $\xi(k)$ to a node for a given $k$. Note
that the number `1' in $i$-th row/\-co\-lumn of the matrix $\mathbf{S}$ gives
the degree of $i$-th node. On the other hand, the mean $\sum_{j=1}^N s(i,j)/N$ of matrix elements in $i$-th row/\-co\-lumn is the average distance $\xi_i$ {\em to
that node}.

The results are averaged over $N_{run}$ independent networks, i.e. various
matrices $\mathbf{S}_{N\times N}$.

\section{Results and discussion}
For the scale-free networks we reproduce $P(k)\propto k^{-\gamma}$ with $\gamma\approx 2.7$, while the theoretical value is 3.0.
The numerical reduction of $\gamma$ is known to be caused by the finite-size effect \cite{ba,drme}.
For the exponential trees the node degree distribution is verified to be $P(k)=2^{-k}$.
The degree distribution for CRG follows the Poisson distribution
$P(k)=\exp(-\langle k\rangle)\cdot\langle k\rangle^k/k!$, with $\langle k\rangle\approx 20$
and $\langle k\rangle\approx 50$ for $p=0.02$ and $p=0.05$, respectively.

\begin{figure}
\begin{center}
\resizebox{0.99\hsize}{!}{\includegraphics{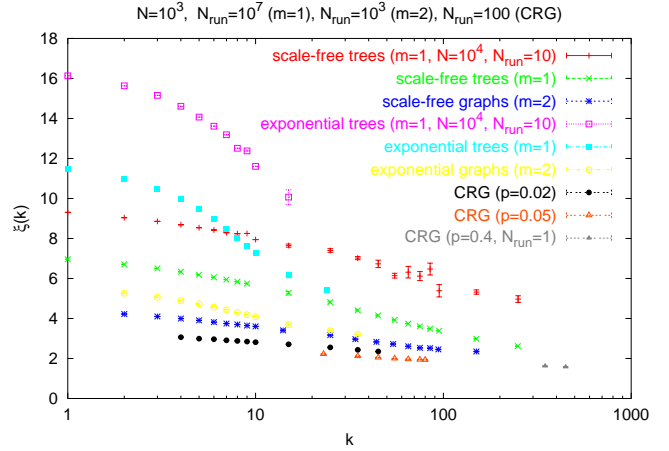}}
\caption{The average distance $\xi(k)$ to a node with given degree $k$ for
different networks.}
\label{fig-xi}
\end{center}
\end{figure}

\begin{figure}
\begin{center}
\resizebox{0.99\hsize}{!}{\includegraphics{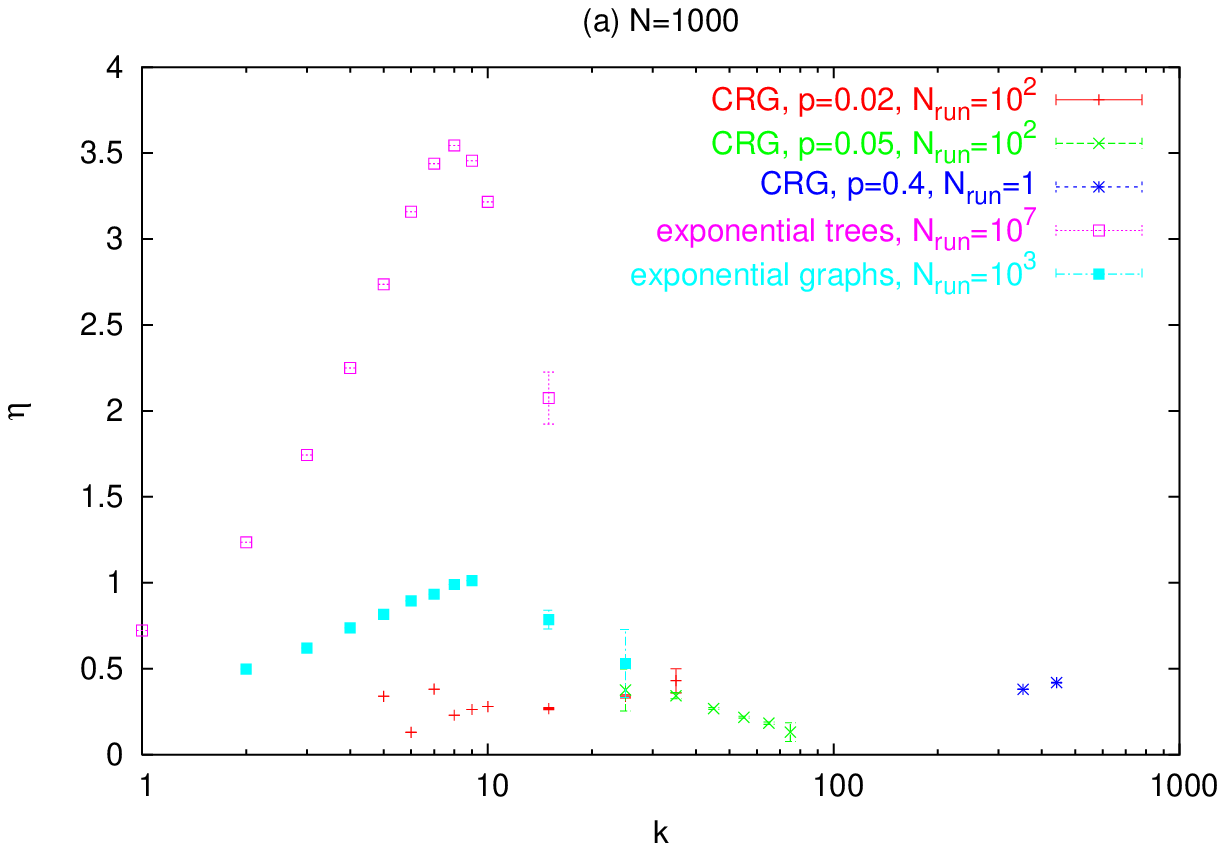}}\\
\resizebox{0.99\hsize}{!}{\includegraphics{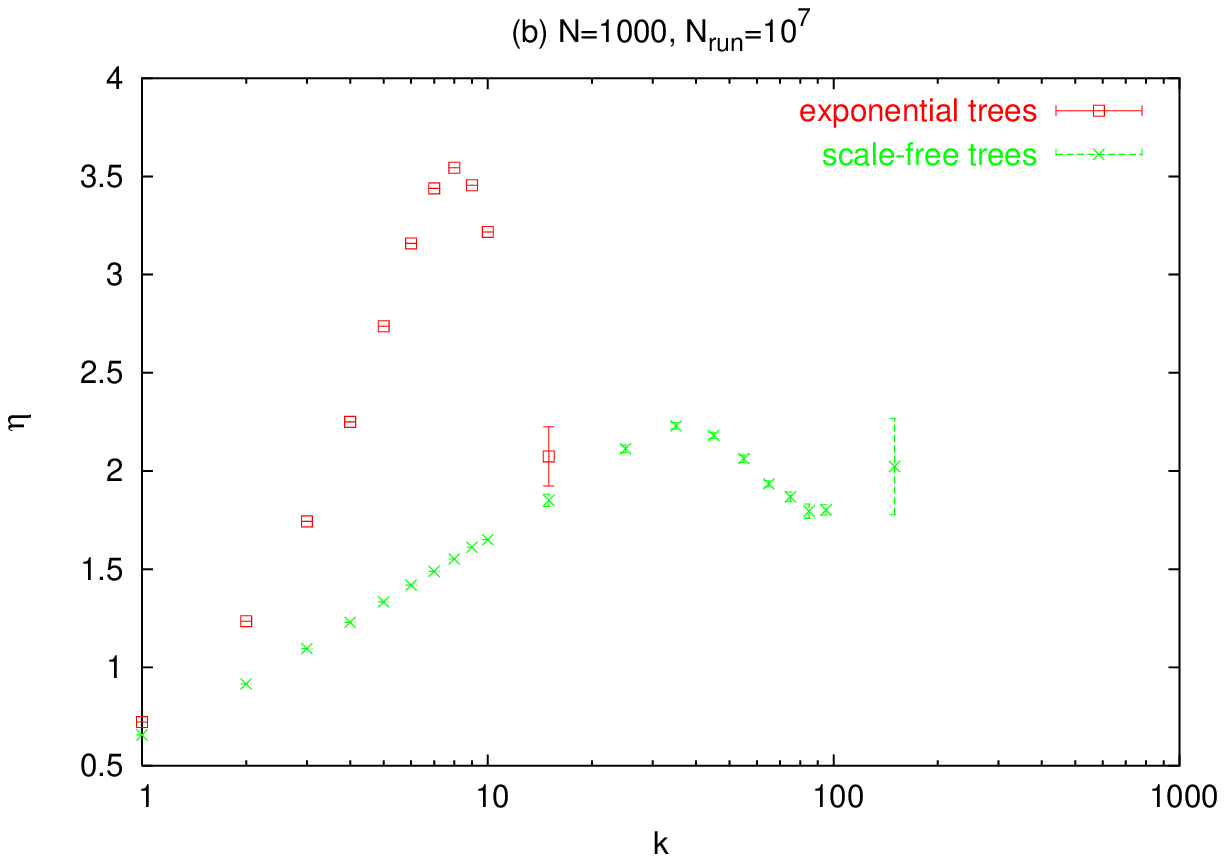}}
\caption{The dependence $\eta(k)$ for
(a) the exponential networks and CRG ($p=0.02$, $0.05$ and $0.4$)
and (b) the exponential trees and the scale-free trees.
The numerical uncertainties are smaller than the symbol size, except the last two bins for trees for which uncertainties are huge and not shown.
} \label{fig-eta}
\end{center}
\end{figure}

In Fig. \ref{fig-xi} the average distance $\xi(k)$ to a node with a given
degree $k$ is presented. Each network contains at least a thousand nodes.
In Fig. \ref{fig-eta} the dependence $\eta(k)$ is shown.
The results are averaged over $N_{run}=10^7$, $10^3$ and $100$ different
networks for trees, simple graphs and CRG, respectively.

As it was explained at the end of the Introduction, the most relevant are the left part of the curves $\eta(k)$, where the degree is small.
In our search, the results for large $k$ reflect the fact that once in our search the nodes of highest possible degree are reached, a further search may not be efficient.
(Note, that in the simulation performed in Ref. \cite{kim}, the search was stopped once the distance from the desired node was one.)

For larger networks, the whole plots presented in Figs. \ref{fig-xi} and \ref{fig-eta} are expected to be stretched toward larger values of $k$.
However, this stretching is logarithmically slow.

In Fig. \ref{fig-xi} the results for trees with $N=10^3$ and $10^4$ are compared.
The main difference is just the shift of the curve upward when $N$ increases.
The slope of the curve, $\eta(k)$, is therefore roughly the same.
The results show that the investigated strategy (MCNS) is most effective for
the exponential trees, where $m=1$. There, the value of the index $\eta $ is
the largest. This is true in particular about
$k=10$, where $\eta$ has a maximum.
The existence of this maximum does not depend on the size $N$ of investigated networks. However, for the exponential networks with
$m=2$, the obtained values of $\xi$ depend much weaker on the degree $k$.
There, the obtained values of $\eta$ are comparable to those of the scale-free
trees. Here again, the size of the network does not influence the
results, but the increase of the number of links $m$ from one to two leads to
a further decrease of the index $\eta $. Finally, for the random graphs the
mean distance $\xi$ practically does not depend on $k$, and the index $\eta$
is close to zero. These conclusions on the scale-free networks and on the
random graphs agree with the results of Ref. \cite{ada}, but
 MCNS applied in an exponential tree is even more effective than in the
scale-free tree.

The explanation of the result is as follows. In the scale-free networks, local
fluctuations of the degree are enhanced by subsequent linkings. In this way,
the structure becomes heterogeneous: multiple centres of high degree can be
created, and the growth concentrates around these centres. This hierarchical
structure of the scale-free networks was described recently in
Ref. \cite{rvb}. Then, MCNS can be misleading, as it always leads to a local
centre; however, sometimes the target is somewhere else. This enhancement is
absent in the exponential networks, and that is why MCNS works better there.
We note that this argumentation works well for trees. For other systems, there
is more than one path between each pair of nodes, and any educated but general
strategy cannot replace the knowledge of where the target is.

Our new tool --- the index $\eta $, defined above --- seems to be useful
for comparing different kinds of networks. In a purely geometrical sense, it
gives the following information: if a node has more edges, how much closer is it to
the network centre, where the mean distance $\xi$ is minimal?
From this point of view, the structure of a given network can be found to be more or less resistant to damage and/or penetration.
This problem is of potential relevance for numerous applications, e.g. in computer science, sociophysics and immunology \cite{stauffer,galam,japan}.

As for our knowledge, the only example of the exponential network is the electrical power grid in western US \cite{valente}.
However, we know examples where the preference of linking is inverted:
new nodes are more likely linked, than old ones. Such  is the case of the
diffusion-limited aggregation, known as DLA, which leads to a formation of
fractal-like dendritic molecules \cite{mea}. If such an anti-preference is
possible, it is sure that some networks also exist where the preference is absent, or at least small.
These latter networks should be close to the exponential ones.
For example, suppose that a network of actresses is investigated, the
preference for old nodes could be weaker.

In conclusion, we have formulated a quantitative criterion for evaluation the
search strategy by linking to a most connected neighbour. We demonstrated
that this strategy is more efficient for the exponential trees than for the
scale-free and random networks.

\begin{acknowledgement}
We are grateful to our anonymous Referees for helpful comments.
The numerical calculations were carried out in ACK\---CY\-F\-RO\-NET\---AGH.
The machine time on SGI~2800 is financed by the Ministry of Science and
Information Technology in Poland under grant No.
KBN/\-SGI2800/\-AGH/\-018/\-2003.
\end{acknowledgement}


\end{document}